\documentclass[prb,twocolumn,preprintnumbers,superscriptaddress]{revtex4}

\usepackage{graphicx,epstopdf}% Include figure files
\usepackage{amssymb,amsmath}
\usepackage{dcolumn}% Align table columns on decimal point
\usepackage{bm}% bold math
\usepackage{color}
\usepackage{enumerate}

\begin{document}

\title{Attractive Coulomb interaction of 2D Rydberg excitons}

\author{V. Shahnazaryan}
\affiliation{Science Institute, University of Iceland, Dunhagi-3, IS-107, Reykjavik, Iceland}
\affiliation{Institute of Mathematics and High Technologies, Russian-Armenian (Slavonic) University, Hovsep Emin 123, 0051, Yerevan, Armenia}
\affiliation{ITMO University, St. Petersburg 197101, Russia}

\author{I. A. Shelykh}
\affiliation{Science Institute, University of Iceland, Dunhagi-3, IS-107, Reykjavik, Iceland}
\affiliation{ITMO University, St. Petersburg 197101, Russia}
\affiliation{Division of Physics and Applied Physics, Nanyang Technological University 637371, Singapore}

\author{O. Kyriienko}
\affiliation{The Niels Bohr Institute, University of Copenhagen, Blegdamsvej 17, DK-2100 Copenhagen, Denmark}

%\date{\today}

\begin{abstract}
We analyze theoretically the Coulomb scattering processes of highly excited excitons in the direct bandgap semiconductor quantum wells. We find that contrary to the interaction of ground state excitons the electron and hole exchange interaction between excited excitons has an attractive character both for $s$- and $p$-type 2D excitons. Moreover, we show that similarly to the three-dimensional (3D) highly excited excitons, the direct interaction of 2D Rydberg excitons exhibits  van der Waals type long-range interaction. The results predict the linear growth of the absolute value of exchange interaction strength with an exciton principal quantum number, and point the way towards enhancement of optical nonlinearity in 2D excitonic systems.
\end{abstract}


% 71.35.-y	Excitons and related phenomena
% 73.21.-b	Electron states and collective excitations in multilayers, quantum wells, mesoscopic, and nanoscale systems
% 73.21.Fg	Quantum wells
% 33.80.Rv	Multiphoton ionization and excitation to highly excited states (e.g., Rydberg states)

\maketitle

\section{Introduction}

The possibility to attain strong and tunable interparticle interactions in a many body system is indispensable for both fundamental studies of strongly correlations and practical exploitation of nonlinear effects. The vast variety of collective effects in cold atom systems \cite{Bloch2008} has profited from usage of Feshbach resonances \cite{Chin2010}. They allow for tunability of s-wave scattering length for atomic collisions, changing the interaction character from a short-range repulsive to an attractive one.
A major step forward in boosting the atomic interaction strength can be performed when atoms are excited to a large principal quantum number Rydberg state \cite{Gallagher1994}. In this case the absolute value of interaction strength grows dramatically, and the interaction potential becomes of long-range nature, leading to the phenomenon of Rydberg blockade \cite{Lukin2001,Gaetan2009,Urban2009}. This facilitates numerous applications in the quantum optics domain \cite{Saffman2010}, where large effective nonlinearity for photons enables efficient photon crystallization \cite{Peyronel2012}, creation of photonic molecules \cite{Firstenberg2013}, ordered pattern formation \cite{Schauss2012} etc.

In the solid state physics, the studies of many body effects and nonlinear quantum optics became possible for the systems of interacting quasiparticles typically probed by light. Here, the prominent examples are indirect excitons \cite{Butov2007,High2012} and exciton polaritons \cite{Carusotto2013}. The latter quasiparticles formed by the microcavity photons and excitons in two dimensional (2D) semiconductor quantum well (QW) are especially valuable for observation of non-equilibrium condensation \cite{Kasprzak2006,Schneider2013}, vortices \cite{Lagoudakis2008}, solitons \cite{Sich2012,Amo2011} and other effects characteristic to weakly nonlinear Bose gas. At the same time, the highly anticipated transition of polaritonics to quantum nonlinear regime is deferred by small and short range exciton-exciton interaction in QWs, which are dominated by \emph{repulsive} Coulomb exchange, while direct interaction contribution is negligible \cite{Ciuti1998,Tassone1999}, except for the narrow energy range where the formation of bipolariton is possible. Therefore, it opens the challenge for system modification to attain strong interaction, or, alternatively, the search for optional strategies which require only weak nonlinearity \cite{Liew2010,Kyriienko2014,Kyriienko2014b}.

Up to date proposals for the enhancement of nonlinearity include hybridization of polaritons with dipolar excitons (dipolaritons) \cite{Cristofolini2012,Christmann2011,Kristinsson2014} and exploitation of the biexcitonic Feshbach resonance \cite{Wouters_2007,Wouters_2014}, though with limited capabilities. A drastic improvement was made in the system of highly excited 3D excitons, being excitonic counterpart of Rydberg atoms physics \cite{Kazimierczuk2014}. There authors reported an observation of the dipolar blockade appearing in bulk Cu$_2$O for giant excitons having principal quantum number up to $n=25$ and $\mu$m diameter. The important consequence of using rather peculiar copper oxide semiconductor is selection rules which allow to optically pump excitons in the $p$-state, where excitons exhibit the long range interaction of dipolar and van der Waals type similarly to Rydberg atoms. At the same time, while results show the potential for strongly nonlinear optics, the requirement of 3D geometry and infeasibility of Cu$_2$O based microcavities hinder its application in the conventional form.

In this article we pose the question of possible achievement of the strong exciton-exciton interaction exploiting highly excited states of excitons in 2D semiconductor quantum wells. We show that for small transferred momenta the interaction of 2D excitons is dominated by short range exchange Coulomb interaction for both $s$ and $p$ exciton types, and find that for excitons with higher than ground principal quantum number ($n>1$) the interaction constant changes the sign, leading to an \emph{attractive} exciton-exciton potential. The absolute value of interaction strength scales linearly with $n$, and increases for small band gap semiconductors. At the same time, similarly to the 3D geometry, the direct interaction of 2D excitons possesses long range nature governed by van der Waals law and grows drastically with $n$. This suggests that 2D Rydberg exciton gas represents the nontrivial system, and can lead to emergence of hybrid repulsive-attractive bosonic mixtures.
%
%The generation of focusing nonlinearity has significant implications for collective polaritonic effects, polarization properties, and can lead to emergence of hybrid repulsive-attractive bosonic mixtures.
% allowing to study repulsive-attractive bosonic mixtures

\section{The model}

The calculation of interaction potential for 2D excitons in the ground state can be done within the Coulomb scattering formalism \cite{Ciuti1998}. The theory can be extended to describe the interaction of excitons in the excited states. The two dimensional exciton wave function with in-plane wave-vector $\mathbf{Q}$ in the general form reads
\begin{equation}
\Psi_{\mathbf{Q},n,m}(\mathbf{r}_e,\mathbf{r}_h)=\frac{1}{\sqrt{A}}\exp[i\mathbf{Q}(\beta_e\mathbf{r}_e+\beta_h\mathbf{r}_h)]\psi_{n,m}(\mathbf{r}_e,\mathbf{r}_h),
\end{equation}
where $\mathbf{r}_e, \mathbf{r}_h$ are in-plane radius vectors of exciton and hole, respectively, and $A$ denotes the normalization area. The coefficients $\beta_e, \beta_h$ are defined as $\beta_{e(h)}=m_{e(h)}/(m_e+m_h)$, where $m_{e(h)}$ is the mass of an electron and a hole, respectively. The internal relative motion is described by \cite{Portnoi}
\begin{align}
\label{eq:psi_nm}
&\psi_{n,m}(\mathbf{r}_e,\mathbf{r}_h)=\frac{1}{2\lambda_{\mathrm{2D}}}\sqrt{\frac{(n-|m|-1)!}{(n-1/2)^3(n+|m|-1)!}}\\ \notag &\hspace{10mm} \left(\frac{|\mathbf{r}_e-\mathbf{r}_h|}{(n-1/2)\lambda_{\mathrm{2D}}} \right)^{|m|} \exp\left[-\frac{|\mathbf{r}_e-\mathbf{r}_h|}{(2n-1)\lambda_{\mathrm{2D}}}\right] \\ \notag &\hspace{12mm} L^{2|m|}_{n-|m|-1}\left[\frac{|\mathbf{r}_e-\mathbf{r}_h|}{(n-1/2)\lambda_{\mathrm{2D}}}\right]\frac{1}{\sqrt{2\pi}}e^{im\varphi},
\end{align}
where $n=1,2,3,\ldots$ is the principal quantum number, $m=0,\pm1,\ldots,\pm n \mp 1$ is the magnetic quantum number, and $\lambda_{\mathrm{2D}}$ is the two-dimensional Bohr radius of the ground state. Here, $L_{n}^{k}[x]$ denotes associated Laguerre polynomial. In the following we consider the narrow quantum well limit and thus disregard exciton motion in the confinement direction.
%%%
\begin{figure}[t!]
    \includegraphics[width=0.75\linewidth]{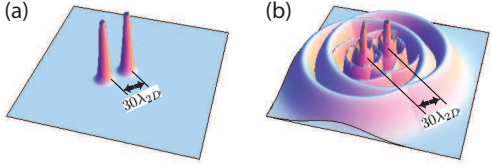}
    \caption{(color online). Real space distribution of exciton envelope wave functions with center-to-center separation distance of $30\lambda_{\mathrm{2D}}$, shown for excitons in (a): 1s state, (b): 6s state.}
    \label{Fig1}
\end{figure}
%%%

Considering excitons with the parallel spin only, the process of Coulomb scattering in reciprocal space associated to the transfer of wave vector $\mathbf{q}$ can be described in the form:
\begin{equation}
(n,m,\mathbf{Q}) + (n',m,\mathbf{Q}') \rightarrow (n,m,\mathbf{Q}+\mathbf{q})+(n',m,\mathbf{Q}'-\mathbf{q}).
\end{equation}
The scattering matrix element consists of four terms:
\begin{equation}
H(n,n',m,\Delta\mathbf{Q},\mathbf{q},\beta_e)=\frac{e^2}{4 \pi \varepsilon \varepsilon_0}\frac{\lambda_{\mathrm{2D}}}{A} I_{\mathrm{tot}}(n,n',m,\Delta\mathbf{Q},\mathbf{q},\beta_e) ,
\label{eq:H}
\end{equation}
where
\begin{align}
&I_{\mathrm{tot}}(n,n',m,\Delta\mathbf{Q},\mathbf{q},\beta_e) \\ \notag
&= I_{\mathrm{dir}}(n,n',m,\mathbf{q},\beta_e) +I^X_{\mathrm{exch}}(n,n',m,\Delta\mathbf{Q},\mathbf{q},\beta_e) \\ \notag
&+I^e_{\mathrm{exch}}(n,n',m,\Delta\mathbf{Q},\mathbf{q},\beta_e) +I^h_{\mathrm{exch}}(n,n',m,\Delta\mathbf{Q},\mathbf{q},\beta_e) .\\ \notag
\label{eq:I}
\end{align}
Here, the first term denotes the direct interaction integral, the second corresponds to the exciton exchange interaction, and two last terms describe an electron and hole exchange integrals (see Appendix A for definitions and details).

Note that in the particular case where the wave vectors and principal quantum numbers of excitons coincide, $\Delta Q = |\mathbf{Q}-\mathbf{Q}'| =0$, and $n=n'$, we have
\begin{align}
&I^X_{\mathrm{exch}}(n,n,m,0,\mathbf{q},\beta_e) = I_{\mathrm{dir}}(n,n,m,\mathbf{q},\beta_e),\\
&I^e_{\mathrm{exch}}(n,n,m,0,\mathbf{q},\beta_e) = I^h_{\mathrm{exch}}(n,n,m,0,\mathbf{q},\beta_e),
\end{align}
and consequently,
\begin{equation}
I_{\mathrm{tot}}(n,m,\mathbf{q},\beta_e)= 2 \big[ I_{\mathrm{dir}}(n,m,\mathbf{q},\beta_e)+I^e_{\mathrm{exch}}(n,m,\mathbf{q},\beta_e)\big].
\label{eq:Ip}
\end{equation}
In the following we are interested in the dependence of interaction on the scattered momentum $\mathbf{q}$, while considering equal exciton center-of-mass momenta, $\Delta Q =0$.

To gain the qualitative understanding of interaction processes for highly excited excitons we shall look at the large $n$ exciton wave function. In particular, Eq. (\ref{eq:psi_nm}) implies that the spatial distribution of exciton drastically increases with principal quantum number. Namely, the higher principal number of excitation is, the larger is a spread of wave function, providing increased overlap between excitons, and consequently leading to the enhanced exciton-exciton interaction. In Fig.~\ref{Fig1}(a) the real space distribution of two excitons in ground state is presented, where the interexciton distance is fixed to $30 \lambda_{\mathrm{2D}}$. The peak-shaped distribution of wavefunctions determines the interaction behavior, which rapidly decreases as distance grows. Panel (b) shows the probability distribution for excitons in $6s$ state, with the same interexciton distance as before (i.e., same density of particles), revealing large overlap of wave functions.

\section{Results}

\subsection{Interaction between $s$-type excitons}

We examined numerically the Coulomb interaction integrals between excitons in $s$ and $p$ states as a function of the scattered momentum $\mathbf{q}$. The calculation was done by multidimensional Monte-Carlo integration with implemented importance sampling algorithm, provided by the numerical integration CUBA library \cite{Hahn_VEGAS}. To be specific, we fixed the electron to exciton mass ratio to the value characteristic to GaAs, $\beta_e = 0.11$. We note that the change of this parameter does not lead to significant quantitative and any qualitative changes of results, and comment on the possible choices for materials subsequently.

\begin{figure}[t]
    \includegraphics[width=1.0\linewidth]{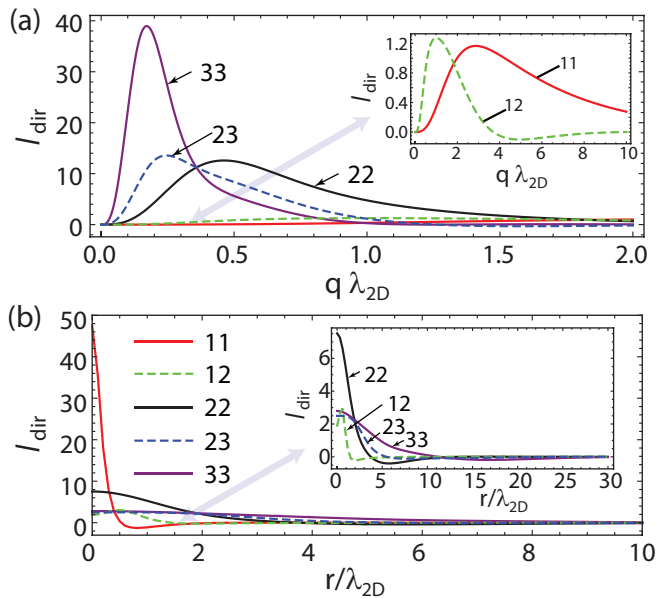}
    \caption{(color online). Interaction of $s$-excitons. (a) Dependence of direct exciton-exciton interaction on the scattered wave vector $q$ (in terms of reverse two-dimensional Bohr radius). Hereafter, the dimensionless value of integration is presented. (b) Real space dependence of direct interaction integral. }
    \label{Fig2}
\end{figure}

We consider the interaction between two $s$-type excitons with the same ($\{n,n'\}=11,22,33$) and different ($\{n,n'\}=12,23$) principal quantum numbers. The results of calculation are plotted in Fig. \ref{Fig2}. Panel (a) shows the direct interaction term as a function of dimensionless transferred momentum for various scattering processes. We find that direct interaction for ground state excitons and excited excitons has the same qualitative behavior, dropping to zero for small $q$ and exhibiting maximum for intermediate momenta. The position of direct interaction peak shifts to smaller transferred wave vectors for increasing $n$, and its magnitude increases radically. We check that the following holds even for very large quantum numbers (up to $n=10$; not shown).
%%%
%%%
In Fig. \ref{Fig2}(b) we plot a 2D Fourier transform of $I_{\mathrm{dir}}[q]$ interaction integral, which represents its real space dependence. The curves depict maximal but finite interaction strength for $n=n'=1$ excitons at a small separation, which rapidly decreases with $r$. For excited states the $r \rightarrow 0$ peak flattens out, while total interaction range increases.

To understand the origin of interaction we examined the large $r$ behavior of the potential for excitons with quantum number in the range $n = 3..10$ (see Appendix B for details). The analysis of interaction tail unveiled the rapid increase of interaction strength with the growth of principal quantum number, being another fingerprint of long range nature of interaction \cite{Gallagher1994,Saffman2010,Kazimierczuk2014}. The corresponding numerical fit of real space interaction dependence revealed the van der Waals nature of potential ($I_{\mathrm{dir}}\propto r^{-6}$), also characteristic to QW excitons in the ground state \cite{Schindler2008}.
%%%
\begin{figure}[t]
    \includegraphics[width=1.0\linewidth]{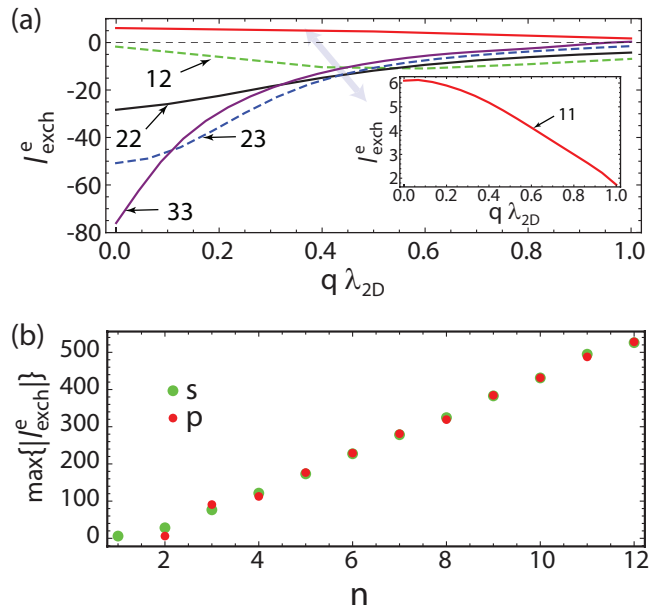}
    \caption{(color online). (a) Dependence of $s$-type exciton-exciton interaction electron exchange term on the scattered wave vector $q$. (b) The maximal absolute value of the exchange interaction plotted as a function of principal quantum number $n$ for $s$ and $p$ states. The linear in $n$ growth of the interaction strength is observed.}
    \label{Fig3}
\end{figure}
%%%

Next, we calculate the Coulomb exchange contribution to $s$-type exciton-exciton interaction. Fig. \ref{Fig3}(a) illustrates the dependence of exchange integral $I_{\mathrm{exch}}$ as a function of $\mathbf{q}$ for different states. For the ground state scattering (inset, curve 11) the interaction is maximal in $q \rightarrow 0$ region, decreasing for large exchanged wave vectors, and has positive sign (repulsive potential). However, already for $n=n'=2$ the sign of exciton interaction changes to \emph{attractive} one, with maximal absolute value at small $q$. The same change applies to higher excited states interaction, and also to cross scattering between ground and excited state excitons. Moreover, we note that the maximal absolute value of potential grows with principal quantum number $n$, representing an enhancement of exchange contribution by increase of effective interaction area due to the spread of wave functions. Consequently, the real space dependence of exchange interaction has the form of an exponential decay, defined by the decrease of wave function overlap area.

Finally, we study the dependence of maximal absolute value of exchange integral as a function of principal quantum number $n=n'$, measured at $q \rightarrow 0$ point. The behavior is shown in Fig. \ref{Fig3}(b) for both $s$ and $p$ excitons, corresponding to a linear increase of the magnitude for large principal quantum numbers, $n>3$, where $s$ and $p$ interaction strengths coincide. At the same time, the clear difference in $\mathrm{max}\{|I_{\mathrm{exch}}^e|\}$ for $s$ and $p$ states is visible at $n\leq 3$ range. This result can be explained by the fact that the radial parts of wave functions of excited states have the same shape at larger radii, although being different at small $r$, relevant for small $n$ excitons.
%%%
\begin{figure}[t]
    \includegraphics[width=1.0\linewidth]{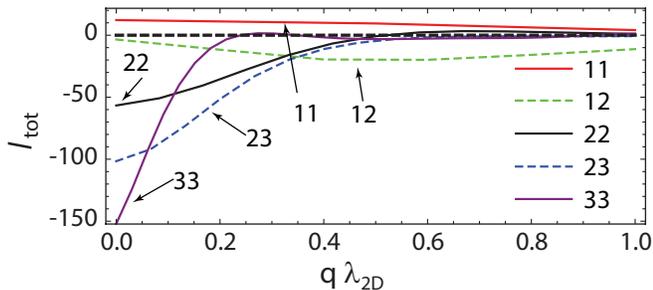}
    \caption{(color online). Overall interaction of $s$-type excitons as a function of the scattered wave vector $q$. For small values of $q$ the interaction is highly attractive due to the dominant exchange interaction, while for large values the direct term prevails, leading to the repulsive character of total interaction. }
    \label{Fig4}
\end{figure}
%%%

The total interaction potential in the case of equal wave vectors and principal quantum numbers, represented by Eq. (\ref{eq:Ip}), is shown in Fig. \ref{Fig4} as a function of the transferred momentum $\textbf{q}$. It reveals that for very small values of $q$ the total interaction for excited states is fully determined by the exchange interaction, being attractive. However, for larger transferred momenta it is replaced by weak repulsion, showing the dominant contribution of the direct interaction term in large $q$ region. Noteworthy, for higher excitation number the region with domination of repulsion is shifted to smaller transferred momenta values. This alternating sign behavior is intriguing as it can potentially lead to the formation of supersolid state \cite{Matuszewski2012}.
%{\color{blue} However, the smallness of the effect drastically decreases its possible experimental observation.}

\subsection{Interaction between $p$-type excitons}

\begin{figure}[t]
    \includegraphics[width=1.0\linewidth]{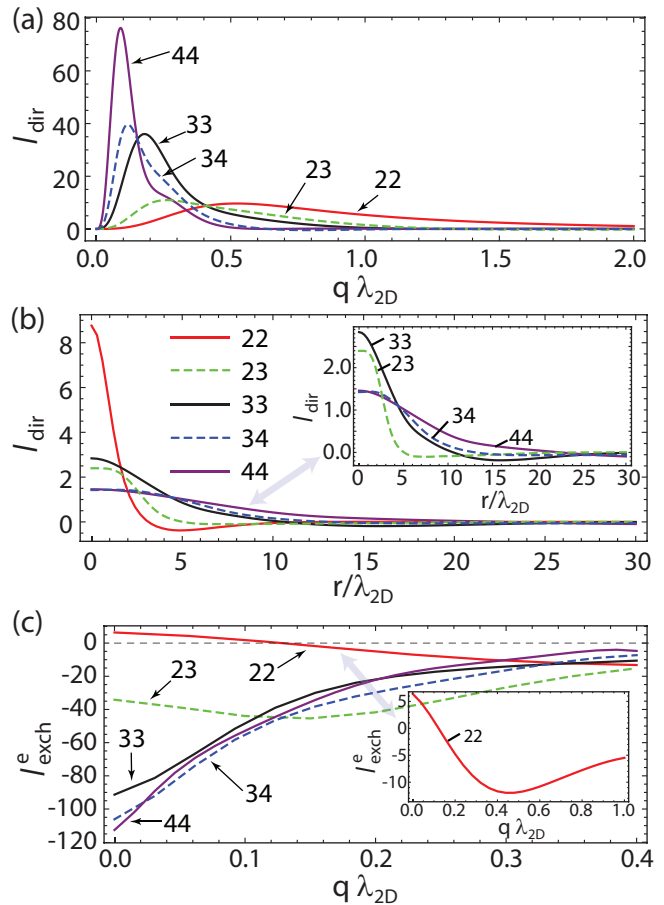}
    \caption{Interaction of $p$-excitons. (a) Dependence of direct exciton-exciton interaction on the scattered wave vector $q$ (in terms of reverse two-dimensional Bohr radius). Here, the dimensionless value of integration is presented. (b) Real space dependence of direct interaction integral. (c) Dependence of electron exchange term on the scattered wave vector $q$. }
    \label{Fig5}
\end{figure}

We proceed with the discussion of direct and exchange Coulomb integrals for two $p$-type excitons. While non-zero angular momentum states are not straightforwardly accessed by optical means in direct-gap  semiconductors (GaAs, GaN, ZnO etc.), one can envisage the situation when these become relevant in the low dimensional structures. As an example they can be created by two photon pumping \cite{Kavokin_2014}. The results of numerical integrations are presented in Fig. \ref{Fig5}. Panel (a) shows the direct interaction integral as a function of transferred momentum for various values of the principal quantum numbers. We note that while qualitatively it has the same behavior as $s$-type excitons, the absolute maxima are higher for $p$-type. The positions of maxima are also shifted to smaller values of transferred momenta. This corresponds to flatter real space dependence of direct interaction, presented in Fig. \ref{Fig5}(b). The effect can be explained by the evidence that for small values of principal quantum number ($n=2,3$) the wave function of $p$-type exciton is much wider than that for $s$-exciton. Noteworthy, while for $p$-shells of 3D excitons the long range interaction has dipole-dipole contribution, it is absent for 2D excitons with non-zero angular momentum.
%%%
%%%

Finally, panel \ref{Fig5}(c) illustrates the exchange term dependence on the transferred wave vector $\textbf{q}$. Analogously to the direct interaction, it has a shape similar to one of $s$-type excitons.

\section{Discussion and outlook}

Previously we have shown that interactions between excited excitonic states in 2D structures have different contributions, which are largely dependent on the main quantum number $n$ and interexciton separation. While very large $n$ excitons physics is expected to be driven by long range interactions, the relevant properties of ground state excitons are defined by short range exchange potential. Thus, we expect the cross-over between to regimes to happen in the range of intermediate $n > 1$, where strong short-range attractive interaction dominates. To increase the overall interaction even further, we consider possible semiconductor materials where Rydberg excitons can be observed. The parameters of 2D Bohr radius, binding energy Ry$_{3D}$, Coulomb interaction prefactor of Eq. (\ref{eq:H}) $\alpha_C \equiv e^2 \lambda_{\mathrm{2D}} /4 \pi\varepsilon\varepsilon_0$, and band gap E$_g$ are collected in Table I for various semiconductors (data is taken from Refs. \cite{Hanada, Singh, Rubio}). One can see that with increasing band gap the exciton Bohr radius decreases, consequently diminishing the interaction constant.
\begin{table}
\centering
    \begin{tabular}{|c|c|c|c|c|}
        \hline
        $ $  & $\lambda_{\mathrm{2D}}~(\mathrm{{\AA}})$ & Ry$_{3D}$ (meV) & $\bm{\alpha_C~(\mu \mathrm{eV} \mu m^2)}$ & E$_g$ (eV) \\ \hline
        InAs & 184.45                 & 1.29           & \textbf{1.75}                         & 0.354      \\ \hline
        GaSb & 111.95                 & 2.05           & \textbf{1.03}                         & 0.726      \\ \hline
        InN  & 36.3                   & 6.47           & \textbf{0.34}                         & 0.78       \\ \hline
        InP  & 46.95                  & 6.13           & \textbf{0.54}                         & 1.344      \\ \hline
        GaAs & 61                     & 4.57           & \textbf{0.68}                         & 1.424      \\ \hline
        CdTe & 30.1                   & 11.70          & \textbf{0.42}                         & 1.5        \\ \hline
        GaN	 & 21.75                  & 17.04          & \textbf{0.32}	                       & 3.2        \\ \hline
        ZnO	 & 10.55	              & 40.22          & \textbf{0.178}	                       & 3.37       \\ \hline
    \end{tabular}
    \caption{The estimation of Bohr radius of two-dimensional exciton, Coulomb interaction constant $\alpha_C \equiv e^2 \lambda_{\mathrm{2D}} /4 \pi\varepsilon\varepsilon_0$, and binding energy (3D) for direct band gap semiconductors. The list is sorted by increasing order of a semiconductor band gap.}
\end{table}
At the same time we note that successful generation of highly excited excitonic states requires large binding energy of excitons, which allows to address separately excitonic states with large $n$.
Therefore, an interplay between interaction strength and exciton energy separation implies mid-bandgap semiconductors (e.g. GaAs) to be relevant for described physics. Additionally, we underline the possible importance of materials with the non-Rydberg excitonic spectrum, represented by transition metal dichalcogenides \cite{Chernikov,Berghauser,Qiu,Berkelbach}, where the described bound can be violated.

Finally, we foresee that optical pumping of $s$ excitonic states does not pose optical selection rules for excitons with different $n$, which allows the creation of $n=1,2,3,...$ exciton mixture. Given its mutually attractive and repulsive interaction, we expect an intriguing collective effects to appear in the system.

\section{Conclusion}

We studied the Coulomb interaction of excited states of excitons in direct gap semiconductors. We show that the exchange interaction of higher states has an attractive character due to the dominant contribution of exchange terms. The linear increase of interaction maxima with increase of the principal quantum number of excitonic state was observed. Contrary to 3D excitons, no dipolar interaction appears for large quantum number 2D excitons, and direct interaction has van der Waals behavior. The results point out an importance of Rydberg excitonic states, and may open the way towards studies of repulsive-attractive bosonic mixtures.\\

\section*{Acknowledgements}The work was supported by the FP7 ITN NOTEDEV network, RISE project 644076 CoExAN, Russian Federal Target Programm, project 14.587.21.0020, and  Singapore Ministry
of Education under AcRF Tier 2 grant MOE2015-T2-1-055. V.S. thanks the Niels Bohr Institute for hospitality.

\appendix
\section{Derivation of matrix elements for Coulomb scattering of Rydberg excitons}

A two-dimensional exciton in $nl$ state with center of mass wave vector $\mathbf{Q}$ is described by the wavefunctions (1) and (2) of the main text, corresponding to center of mass and internal motions, respectively.
The spin degree of freedom can be introduced in the following way. The total angular momentum projection of conduction electron on the growth axis is $s_e=\pm 1/2$. In the current work we restrict ourselves with the consideration of heavy-hole exctions. The angular momentum projection of heavy holes is $j_h=\pm 3/2$. Correspondingly, we have four independent heavy-hole exciton states: the dipole-active states $|J_z=\pm1\rangle=|s_e=\mp 1/2, j_h=\pm 3/2\rangle$, and the dark states $|J_z=\pm2\rangle=|s_e=\pm 1/2, j_h=\pm 3/2\rangle$. Further, in a general case exciton state with total momentum $|S\rangle$ can be defined as $\chi_S(s_e,j_h)=\langle s_e, j_h|S\rangle$ (see e.g. Ref. \cite{Ciuti1998} for a detailed description).

We proceed considering the Coloumb scattering of excitons. We are interested in the processes of elastic scattering which conserve total spin and principal quantum numbers of excitons. They correspond to scattering processed described as:
\begin{equation}
(nl,\mathbf{Q},S)+(n'l,\mathbf{Q}',S)\rightarrow(nl,\mathbf{Q}+\mathbf{q},S)+(n'l,\mathbf{Q}'-\mathbf{q},S),
\end{equation}
where we defined a distinct exciton spin state  $|S\rangle= | s, j\rangle$, yielding $\chi_S(s_e,j_h)=\langle s_e, j_h|S\rangle=\delta_{s_e,s}\delta_{j_h,j}$.

Within the Hartree-Fock approximation, the two-exciton initial state having the same spin is described by the following wave function:
\begin{widetext}
\begin{equation}
    \begin{aligned}
&\Phi_{\mathbf{Q}\mathbf{Q}'nn'}^{S}(\mathbf{r}_e,s_e,\mathbf{r}_h,j_h,\mathbf{r}_{e'},s_{e'},\mathbf{r}_{h'},j_{h'})=\frac{1}{\sqrt{2}}\left\{\frac{1}{\sqrt{2}}
\left[\Psi_{\mathbf{Q},n}(\mathbf{r}_e,\mathbf{r}_h)\chi_S(s_e,s_h) \Psi_{\mathbf{Q}',n'}(\mathbf{r}_{e'},\mathbf{r}_{h'})\chi_{S}(s_{e'},s_{h'}) \right. \right. \\
&\left. +\Psi_{\mathbf{Q},n}(\mathbf{r}_{e'}, \mathbf{r}_{h'})\chi_S(s_{e'},s_{h'})\Psi_{\mathbf{Q}', n'}(\mathbf{r}_e,\mathbf{r}_h)\chi_{S}(s_e,s_h)\right]-\frac{1}{\sqrt{2}}\left[\Psi_{\mathbf{Q},n}(\mathbf{r}_{e'},\mathbf{r}_h)\chi_S(s_{e'},s_h) \Psi_{\mathbf{Q}',n'}(\mathbf{r}_e,\mathbf{r}_{h'})\chi_{S}(s_e,s_{h'}) \right.\\
&+\left.\left. \Psi_{\mathbf{Q},n}(\mathbf{r}_e,\mathbf{r}_{h'})\chi_S(s_e,s_{h'})\Psi_{\mathbf{Q}',n'}(\mathbf{r}_{e'},\mathbf{r}_h)\chi_{S}(s_{e'},s_h)\right]\right\}=\delta_{s_e,s}\delta_{s_{e'},s} \delta_{j_h,j}\delta_{j_{h'},j} \left\{\frac{1}{2}\left[ \Psi_{\mathbf{Q},n}(\mathbf{r}_e,\mathbf{r}_h) \Psi_{\mathbf{Q}',n'}(\mathbf{r}_{e'},\mathbf{r}_{h'}) \right. \right.\\
&+\left.\left. \Psi_{\mathbf{Q},n}(\mathbf{r}_{e'}, \mathbf{r}_{h'})\Psi_{\mathbf{Q}', n'}(\mathbf{r}_e,\mathbf{r}_h) \right]- \frac{1}{2}\left[ \Psi_{\mathbf{Q},n}(\mathbf{r}_{e'},\mathbf{r}_h)\Psi_{\mathbf{Q}',n'}(\mathbf{r}_e,\mathbf{r}_{h'})+ \Psi_{\mathbf{Q},n}(\mathbf{r}_e,\mathbf{r}_{h'})\Psi_{\mathbf{Q}',n'}(\mathbf{r}_{e'},\mathbf{r}_h) \right] \right\}.\\
    \end{aligned}
\end{equation}
\end{widetext}
The Hamiltonian can be written in the form:
\begin{equation}
\hat{H}=\hat{H}_1(\mathbf{r}_e,\mathbf{r}_h)+\hat{H}_2(\mathbf{r}_{e'},\mathbf{r}_{h'})+V_{int}(\mathbf{r}_e,\mathbf{r}_h,\mathbf{r}_{e'},\mathbf{r}_{h'}),
\end{equation}
where $\hat{H}_j$ corresponds to the energy of $j-$th exciton, and $V_{int}$ denotes the Coulomb interaction potential between particles.

The intra-exciton terms read
\begin{equation}
\hat{H}_1(\mathbf{r}_e,\mathbf{r}_h)=-\frac{\hbar^2}{2m_e}\Delta_e-\frac{\hbar^2}{2m_h}\Delta_h-V(|\mathbf{r}_e-\mathbf{r}_h|),
\end{equation}
\begin{equation}
\hat{H}_2(\mathbf{r}_{e'},\mathbf{r}_{h'})=-\frac{\hbar^2}{2m_e}\Delta_{e'}-\frac{\hbar^2}{2m_h}\Delta_{h'}-V(|\mathbf{r}_{e'}-\mathbf{r}_{h'}|),
\end{equation}
and each consists of kinetic and potential energy contributions. $V(r)=\frac{e^2}{4 \pi \varepsilon_0 \varepsilon r}$ corresponds to the Coulomb interaction energy, screened by the static dielectric constant $\varepsilon$; $\varepsilon_0$ is the vacuum permittivity.

The inter-exciton interaction part can be written as
\begin{equation}
    \begin{aligned}
V_{int}(\mathbf{r}_e,\mathbf{r}_h,\mathbf{r}_{e'},\mathbf{r}_{h'})=&-V(|\mathbf{r}_e-\mathbf{r}_{h'}|)-V(|\mathbf{r}_{e'}-\mathbf{r}_h|)\\
                                                                   &+V(|\mathbf{r}_e-\mathbf{r}_{e'}|)+V(|\mathbf{r}_h-\mathbf{r}_{h'}|),\\
    \end{aligned}
\end{equation}
where four possible interactions are accounted.
The scattering amplitude of the process described by Eq. (3) of main text is given by the matrix element:
%%%
\begin{figure*}
    \includegraphics[width=0.8\linewidth]{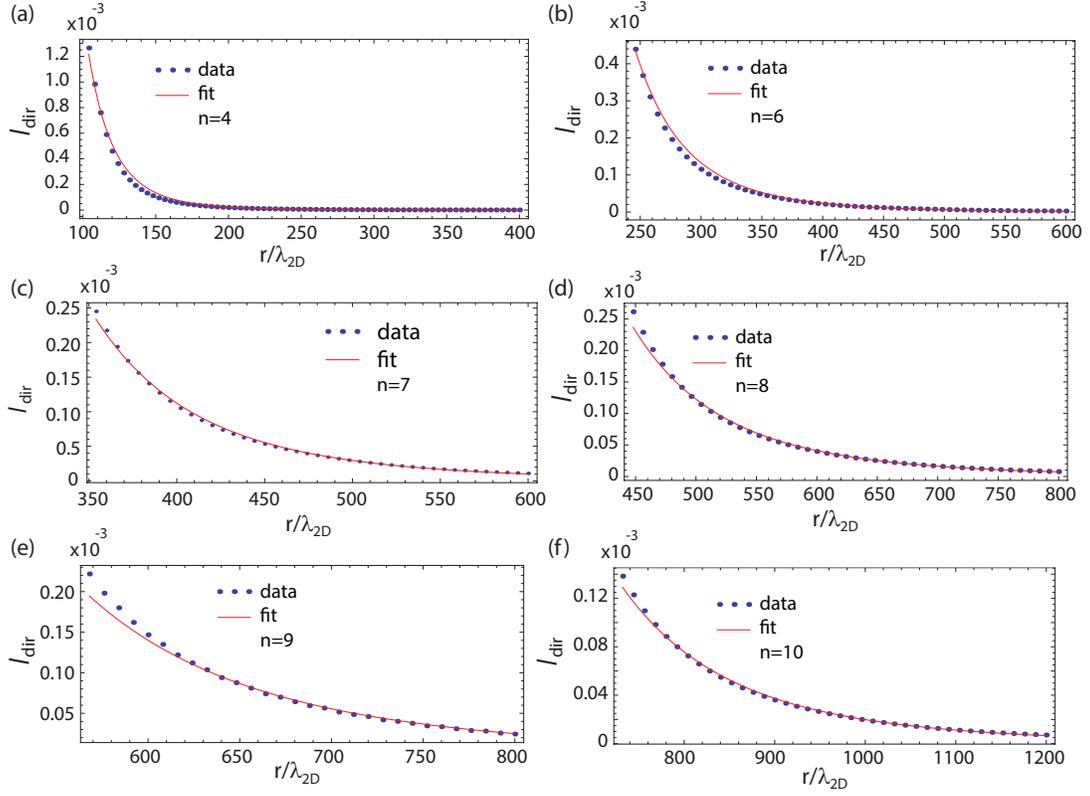}
    \caption{Dependence of long range direct exciton-exciton interaction on the separation distance $r$. The panels (a)-(f) correspond to the interaction of states with principal quantum number $n=4, 6, 7, 8, 9, 10$, respectively. For each value of $n$ the numerical fit shows $\propto r^{-6}$ dependence, characteristic to van der Waals interaction. }
    \label{Fig6}
\end{figure*}
%%%

\begin{widetext}
\begin{equation}
    \begin{aligned}
&H_{nn'mS}(\mathbf{Q},\mathbf{Q}',\mathbf{q})=\int d^2 \mathbf{r}_e \sum_{s_e} \int d^2 \mathbf{r}_h \sum_{j_h} \int d^2 \mathbf{r}_{e'} \sum_{s_{e'}} \int d^2 \mathbf{r}_{h'} \sum_{j_{h'}}\\
&\Phi_{\mathbf{Q}\mathbf{Q}'nn'}^{*S}(\mathbf{r}_e,s_e,\mathbf{r}_h,j_h,\mathbf{r}_{e'},s_{e'},\mathbf{r}_{h'},j_{h'}) V_{int}(\mathbf{r}_e,\mathbf{r}_h,\mathbf{r}_{e'},\mathbf{r}_{h'}) \Phi_{\mathbf{Q}+\mathbf{q}\mathbf{Q}'-\mathbf{q}nn'}^{S}(\mathbf{r}_e,s_e,\mathbf{r}_h,j_h,\mathbf{r}_{e'},s_{e'},\mathbf{r}_{h'},j_{h'})\\
&=\frac14 \delta_{s_e,s}\delta_{s_{e'},s} \delta_{j_h,j}\delta_{j_{h'},j} \\
&\left[4\int d^2\mathbf{r}_ed^2 \mathbf{r}_hd^2\mathbf{r}_{e'}d^2\mathbf{r}_{h'}\Psi^*_{\mathbf{Q},n}(\mathbf{r}_e,\mathbf{r}_h)\Psi^*_{\mathbf{Q}',n'}(\mathbf{r}_{e'},\mathbf{r}_{h'}) V_{int}(\mathbf{r}_e,\mathbf{r}_h,\mathbf{r}_{e'},\mathbf{r}_{h'}) \Psi_{\mathbf{Q}+\mathbf{q},n}(\mathbf{r}_e,\mathbf{r}_h)\Psi_{\mathbf{Q}'-\mathbf{q},n'}(\mathbf{r}_{e'},\mathbf{r}_{h'}) \right. \\
&+4\int d^2\mathbf{r}_ed^2 \mathbf{r}_hd^2\mathbf{r}_{e'}d^2\mathbf{r}_{h'}\Psi^*_{\mathbf{Q},n}(\mathbf{r}_e,\mathbf{r}_h)\Psi^*_{\mathbf{Q}',n'}(\mathbf{r}_{e'},\mathbf{r}_{h'}) V_{int}(\mathbf{r}_e,\mathbf{r}_h,\mathbf{r}_{e'},\mathbf{r}_{h'}) \Psi_{\mathbf{Q}+\mathbf{q},n}(\mathbf{r}_{e'},\mathbf{r}_{h'})\Psi_{\mathbf{Q}'-\mathbf{q},n'}(\mathbf{r}_e,\mathbf{r}_h) \\
&-4\int d^2\mathbf{r}_ed^2 \mathbf{r}_hd^2\mathbf{r}_{e'}d^2\mathbf{r}_{h'}\Psi^*_{\mathbf{Q},n}(\mathbf{r}_e,\mathbf{r}_h)\Psi^*_{\mathbf{Q}',n'}(\mathbf{r}_{e'},\mathbf{r}_{h'}) V_{int}(\mathbf{r}_e,\mathbf{r}_h,\mathbf{r}_{e'},\mathbf{r}_{h'}) \Psi_{\mathbf{Q}+\mathbf{q},n}(\mathbf{r}_e,\mathbf{r}_{h'})\Psi_{\mathbf{Q}'-\mathbf{q},n'}(\mathbf{r}_{e'},\mathbf{r}_h) \qquad \\
&-4\left. \int d^2\mathbf{r}_ed^2 \mathbf{r}_hd^2\mathbf{r}_{e'}d^2\mathbf{r}_{h'}\Psi^*_{\mathbf{Q},n}(\mathbf{r}_e,\mathbf{r}_h)\Psi^*_{\mathbf{Q}',n'}(\mathbf{r}_{e'},\mathbf{r}_{h'}) V_{int}(\mathbf{r}_e,\mathbf{r}_h,\mathbf{r}_{e'},\mathbf{r}_{h'}) \Psi_{\mathbf{Q}+\mathbf{q},n}(\mathbf{r}_{e'},\mathbf{r}_h)\Psi_{\mathbf{Q}'-\mathbf{q},n'}(\mathbf{r}_e,\mathbf{r}_{h'})\right] \\
&= \delta_{s_e,s}\delta_{s_{e'},s} \delta_{j_h,j}\delta_{j_{h'},j} \left[H_{\mathrm{dir}}(n,n',\mathbf{Q},\mathbf{Q}',\mathbf{q})+ H^X_{\mathrm{exch}}(n,n',\mathbf{Q},\mathbf{Q}',\mathbf{q})+H^e_{\mathrm{exch}}(n,n',\mathbf{Q},\mathbf{Q}',\mathbf{q})+H^h_{\mathrm{exch}}(n,n',\mathbf{Q},\mathbf{Q}',\mathbf{q})\right],\\
    \end{aligned}
\end{equation}
where four terms correspond to direct interaction, exciton exchange, electron exchange, and hole exchange. They can be written explicitly as:
\begin{equation}
    \begin{aligned}
&H_{\mathrm{dir}}(n,n',m,\mathbf{q})=\frac{\alpha_C}{A} I_{\mathrm{dir}}(n,n',m,\mathbf{q})=\frac{\alpha_C}{A} \frac{(n-|m|-1)!(n'-|m|-1)!}{2^4\pi^2(n-1/2)^3(n'-1/2)^3(n+|m|-1)!(n'+|m|-1)!} \frac{(2\pi)^3}{\lambda_{\mathrm{2D}}q} \\ &\int\limits_0^{\infty} \int\limits_0^{\infty} \left[ -J_0(\beta_h \lambda_{\mathrm{2D}}q x)J_0(\beta_e \lambda_{\mathrm{2D}}q x')-J_0(\beta_e \lambda_{\mathrm{2D}}q x)J_0(\beta_h \lambda_{\mathrm{2D}}q x')+J_0(\beta_h \lambda_{\mathrm{2D}}q x)J_0(\beta_h \lambda_{\mathrm{2D}}q x')+J_0(\beta_e \lambda_{\mathrm{2D}}q x)J_0(\beta_e \lambda_{\mathrm{2D}}q x') \right] \\
&\left[\frac{x}{n-1/2}\right]^2 e^{-\frac{x}{n-1/2}} \left[L^{2|m|}_{n-|m|-1}\left(\frac{x}{n-1/2}\right)\right]^2 x dx \left[\frac{x'}{n'-1/2}\right]^2 e^{-\frac{x'}{n'-1/2}} \left[L^{2|m|}_{n'-|m|-1}\left(\frac{x'}{n'-1/2}\right)\right]^2 x' dx', \\
    \end{aligned}
\label{dir}
\end{equation}
\begin{equation}
    \begin{aligned}
&H^X_{\mathrm{exch}}(n,n',m,\Delta\mathbf{Q},\mathbf{q})=\frac{\alpha_C}{A} I^X_{\mathrm{exch}}(n,n',m,\Delta\mathbf{Q},\mathbf{q}) =\frac{\alpha_C}{A} \frac{(n-|m|-1)!(n'-|m|-1)!}{2^4\pi^2(n-1/2)^3(n'-1/2)^3(n+|m|-1)!(n'+|m|-1)!} \\
&\int d^2\mathbf{x} d^2\mathbf{y}_1 d^2 \mathbf{y}_2 e^{i(\Delta\mathbf{Q}-\mathbf{q})\lambda_{\mathrm{2D}}\left[\beta_e\mathbf{y}_1+\beta_h\mathbf{y}_2+(\beta_h-\beta_e)\mathbf{x}\right]} \frac{x^2|\mathbf{y}_2-\mathbf{y}_1-\mathbf{x}|^2}{(n-1/2)^2(n'-1/2)^2} e^{-(x+|\mathbf{y}_2-\mathbf{y}_1-\mathbf{x}|)\left[\frac{1}{n-1/2}+\frac{1}{n'-1/2}\right]} L^{2|m|}_{n-|m|-1}\left(\frac{x}{n-1/2}\right) \\
&L^{2|m|}_{n'-|m|-1}\left(\frac{|\mathbf{y}_2-\mathbf{y}_1-\mathbf{x}|}{n'-1/2}\right) L^{2|m|}_{n-|m|-1}\left(\frac{|\mathbf{y}_2-\mathbf{y}_1-\mathbf{x}|}{n-1/2}\right) L^{2|m|}_{n'-|m|-1}\left(\frac{x}{n'-1/2}\right) \left[ -\frac{1}{y_1} -\frac{1}{y_2} +\frac{1}{|\mathbf{y}_1+\mathbf{x}|} +\frac{1}{|\mathbf{y}_2-\mathbf{x}|} \right], \\
    \end{aligned}
\label{X_exch}
\end{equation}
\begin{equation}
    \begin{aligned}
&H^e_{\mathrm{exch}}(n,n',m,\Delta\mathbf{Q},\mathbf{q})=\frac{\alpha_C}{A} I^e_{\mathrm{exch}}(n,n',m,\Delta\mathbf{Q},\mathbf{q})=-\frac{\alpha_C}{A} \frac{(n-|m|-1)!(n'-|m|-1)!}{2^4\pi^2(n-1/2)^3(n'-1/2)^3(n+|m|-1)!(n'+|m|-1)!}  \int d^2\mathbf{x} \\
& d^2 \mathbf{y}_1 d^2\mathbf{y}_2 e^{i\beta_e\lambda_{\mathrm{2D}}\Delta\mathbf{Q}(\mathbf{y}_1+\mathbf{x})} e^{i\lambda_{\mathrm{2D}}\mathbf{q}\left[\beta_h\mathbf{y}_2-\beta_e\mathbf{y}_1-\mathbf{x}\right]}
\left[\frac{x}{n-1/2}\frac{|\mathbf{y}_2-\mathbf{y}_1-\mathbf{x}|}{n'-1/2}\frac{y_1}{n-1/2}\frac{y_2}{n'-1/2}\right]^{|m|} e^{-\frac{x}{2n-1}} e^{-\frac{|\mathbf{y}_2-\mathbf{y}_1-\mathbf{x}|}{2n'-1}} e^{-\frac{y_1}{2n-1}} e^{-\frac{y_2}{2n'-1}}\\
&L^{2|m|}_{n-|m|-1}\left(\frac{x}{n-\frac12}\right) L^{2|m|}_{n'-|m|-1}\left(\frac{|\mathbf{y}_2-\mathbf{y}_1-\mathbf{x}|}{n'-\frac12}\right) L^{2|m|}_{n-|m|-1}\left(\frac{y_1}{n-\frac12}\right) L^{2|m|}_{n'-|m|-1}\left(\frac{y_2}{n'-\frac12}\right) \left[ -\frac{1}{y_2} -\frac{1}{y_1} +\frac{1}{|\mathbf{y}_1+\mathbf{x}|} +\frac{1}{|\mathbf{y}_2-\mathbf{x}|} \right],\\
    \end{aligned}
\label{e_exch}
\end{equation}
\begin{equation}
    \begin{aligned}
&H^h_{\mathrm{exch}}(n,n',m,\Delta\mathbf{Q},\mathbf{q})= \frac{\alpha_C}{A} I^h_{\mathrm{exch}}(n,n',m,\Delta\mathbf{Q},\mathbf{q}) =-\frac{\alpha_C}{A} \frac{(n-|m|-1)!(n'-|m|-1)!}{2^4\pi^2(n-1/2)^3(n'-1/2)^3(n+|m|-1)!(n'+|m|-1)!} \int d^2\mathbf{x} \\
& d^2 \mathbf{y}_1 d^2\mathbf{y}_2 e^{i\beta_e\lambda_{\mathrm{2D}}\Delta\mathbf{Q}(\mathbf{y}_2-\mathbf{x})} e^{i\lambda_{\mathrm{2D}}\mathbf{q}\left[-\beta_h\mathbf{y}_2+\beta_e\mathbf{y}_1+\mathbf{x}\right]}
\left[\frac{x}{n-1/2}\frac{|\mathbf{y}_2-\mathbf{y}_1-\mathbf{x}|}{n'-1/2}\frac{y_1}{n-1/2}\frac{y_2}{n'-1/2}\right]^{|m|} e^{-\frac{x}{2n-1}} e^{-\frac{|\mathbf{y}_2-\mathbf{y}_1-\mathbf{x}|}{2n'-1}} e^{-\frac{y_1}{2n-1}} e^{-\frac{y_2}{2n'-1}}\\
&L^{2|m|}_{n-|m|-1}\left(\frac{x}{n-\frac12}\right) L^{2|m|}_{n'-|m|-1}\left(\frac{|\mathbf{y}_2-\mathbf{y}_1-\mathbf{x}|}{n'-\frac12}\right) L^{2|m|}_{n-|m|-1}\left(\frac{y_1}{n-\frac12}\right) L^{2|m|}_{n'-|m|-1}\left(\frac{y_2}{n'-\frac12}\right) \left[ -\frac{1}{y_2} -\frac{1}{y_1} +\frac{1}{|\mathbf{y}_1+\mathbf{x}|} +\frac{1}{|\mathbf{y}_2-\mathbf{x}|} \right],\\
    \end{aligned}
\label{h_exch}
\end{equation}
\end{widetext}
where we defined $\alpha_C \equiv e^2 \lambda_{\mathrm{2D}} /4 \pi\varepsilon\varepsilon_0$.

During the derivation the following radius vector transformations were used: $\mathbf{\rho}=\mathbf{r}_e-\mathbf{r}_h$, $\mathbf{R}=\beta_e\mathbf{r}_e+\beta_h\mathbf{r}_h$, $\mathbf{\rho}'=\mathbf{r}_{e'}-\mathbf{r}_{h'}$, $\mathbf{R}'=\beta_e\mathbf{r}_{e'}+\beta_h\mathbf{r}_{h'}$, $\mathbf{\xi}=\mathbf{R}-\mathbf{R}'$, $\mathbf{\sigma}=\frac{\mathbf{R}+\mathbf{R}'}{2}$, $\Delta\mathbf{Q}=\mathbf{Q}'-\mathbf{Q}$, $\mathbf{x}=\frac{\mathbf{\rho}}{\lambda_{\mathrm{2D}}}$, $\mathbf{x}'=\frac{\mathbf{\rho}'}{\lambda_{\mathrm{2D}}}$, $\mathbf{y}_1=\frac{\mathbf{\xi}-\beta_e\mathbf{\rho}-\beta_h\mathbf{\rho}'}{\lambda_{\mathrm{2D}}}$, $\mathbf{y}_2=\frac{\mathbf{\xi}+\beta_h\mathbf{\rho}+\beta_e\mathbf{\rho}'}{\lambda_{\mathrm{2D}}}$.\\

\section{Long range interaction}

The current section is devoted to the detailed study of the long range behavior of exciton-exciton direct interaction. For different values of principal quantum number $n$ we examine the real space dependence at very large separation distances. The results are presented in Fig. \ref{Fig6}. The numerical fits show $I_{\mathrm{dir}} \propto r^{-6}$ dependence for interactions of excitons with different quantum numbers $n$. This confirms the van der Waals long range behavor of interaction potential. As it is expected, the distance where van der Waals behavior becomes relevant rapidly grows with the increase of the principal quantum number.
%%%
\begin{figure}[h!]
    \includegraphics[width=1.0\linewidth]{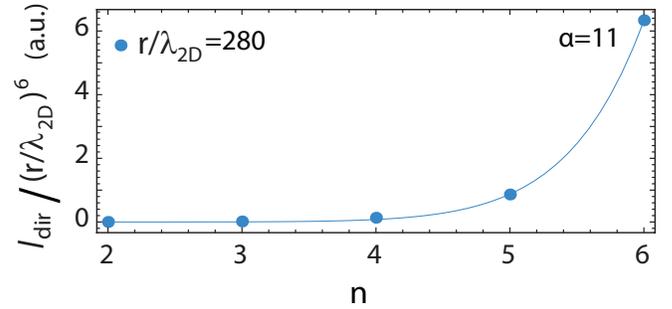}
    \caption{Dependence of direct Coulomb interaction on the excitation number $n$ at different fixed separation distances. The numerical fit shows $\propto n^{\alpha}$ dependence.}
    \label{Fig7}
\end{figure}
%%%

The characteristic feature of van der Waals interaction is the power dependence on the excitation number. To check this, we examined the dependence of the direct interaction strength on the excitation number for different fixed values of the separation distance. The sample of results is presented in Fig. \ref{Fig7}, where the power dependence $\propto n^{\alpha}$ is clearly seen. We observe that due to small number of points the power $\alpha$ can lie in the 7 to 12 range, and expect it to be equal $\alpha=11$ if large $n$ are considered.

%%%%%%%%%%%%%%%%%%%%%%%%%%%%%%%%%

\end{document}